\documentclass[
    ,final            
  ,numberedheadings 
  ]
  {aipproc}

\usepackage{theorem,amsmath,amssymb,mathrsfs}
\usepackage{float,supertabular}
\usepackage{eucal}

\layoutstyle{6x9}

\def\index#1{}
\def\<{\langle}\def\>{\rangle}
\def\:{\hbox{\bf :}}
\renewcommand{\geq}{\geqslant}\renewcommand{\leq}{\leqslant}

\def\Reals{\mathbb R}\def\Cmplx{\mathbb C}
\def\map#1{{\mathscr{#1}}}

\def\set#1{{\sf #1}}\def\alg#1{{\mathcal #1}}\def\aA{\alg{A}}
\def\aI{\alg{I}}

\def\Op#1{\operatorname{Op}_{#1}}

\def\dim{\operatorname{dim}}\def\adm{\operatorname{dim}}
\def\idim#1{\operatorname{dim}_\#(#1)}

\def\sH{\set{H}}\def\sW{\set{W}}

\def\qed{$\,\blacksquare$\par}
\def\eg{e. g. }\def\ie{i. e. }
\def\n#1{|\!|#1|\!|}
\def\dag{\dagger}\def\eff{{\rm eff}}\def\dyn{{\rm dyn}}
 
\newtheorem{definition}{Definition}

\newtheorem{corollary}{Corollary}
\newtheorem{theorem}{Theorem}
\newtheorem{gaxiom}{General Axiom}
\newtheorem{postulate}{Postulate}
\newtheorem{grule}{Rule}
\def\Proof{\medskip\par\noindent{\bf Proof. }}
\theoremstyle{remark}\newtheorem{remark}{{\bf Remark}}
\theoremstyle{example}

\def\trnsfrm#1{\mathscr #1}
\def\tA{\trnsfrm A}\def\tB{\trnsfrm B}\def\tC{\trnsfrm C}\def\tD{\trnsfrm D}
\def\tS{\trnsfrm S}\def\tG{\trnsfrm G}
\def\tI{\trnsfrm I}\def\tT{\trnsfrm T}\def\tU{\trnsfrm U}\def\tX{\trnsfrm X}

\def\AA{\mathbb A}\def\AB{\mathbb B}\def\AC{\mathbb C}\def\AL{\mathbb L}
\def\cA{{\underline{\tA}}}\def\cB{\underline{\tB}}\def\cC{\underline{\tC}}\def\cT{\underline{\tT}}
\def\cI{{\underline{\tI}}}
\def\cX{{\underline{\tX}}}

\def\Stset{{\mathfrak S}}\def\Wset{{\mathfrak W}}
\def\Trnset{{\mathfrak T}}
\def\Cntset{{\mathfrak P}}
\def\Idx#1{%
}

\begin{document}
\title{Operational Axioms for Quantum Mechanics
\footnote{Work presented at the conference {\em 
Foundations of Probability and Physics-4, Quantum Theory: Reconsideration of Foundations-3} 
held on 4-9 June at the International Centre for Mathematical Modeling  in Physics, Engineering and
Cognitive Sciences,  V\"axj\"o University,  Sweden.}
}  
\author{Giacomo Mauro D'Ariano}{ address={{\em QUIT} Group,
    Dipartimento di Fisica ``A. Volta'', via Bassi 6,
    I-27100 Pavia, Italy, {\em http://www.qubit.it}\\
    Department of Electrical and Computer Engineering, Northwestern University, Evanston, IL 60208}}
\begin{abstract} The mathematical formulation of Quantum Mechanics in terms of complex Hilbert space
  is derived for finite dimensions, starting from a general definition of {\em physical experiment}
  and from five simple Postulates concerning {\em experimental accessibility and simplicity}. For
  the infinite dimensional case, on the other hand, a C${}^*$-algebra representation of physical
  transformations is derived, starting from just four of the five Postulates via a
  Gelfand-Naimark-Segal (GNS) construction. The present paper simplifies and sharpens the previous
  derivation in Ref.  \cite{dariano-losini2005}.  The main ingredient of the axiomatization is the
  postulated existence of {\em faithful states} that allows one to calibrate the experimental
  apparatus. Such notion is at the basis of the operational definitions of the scalar product and of
  the {\em transposed} of a physical transformation. What is new in the present paper with respect
  to Ref.  \cite{dariano-losini2005}, is the operational deduction of an involution corresponding to
  the {\em complex-conjugation} for effects, whose extension to transformations allows to define the
  {\em adjoint} of a transformation when the extension is composition-preserving. The existence of
  such composition-preserving extension among possible extensions is analyzed.
\end{abstract}
\maketitle
\section{Introduction}
Quantum Mechanics has been universally accepted as a general law of nature that applies to the
entire physical domain, at any size and energy, and no experiment whatsoever has shown the slightest
deviation from what the theory predict. However, regardless such unprecedented predicting power, the
theory leaves us with a distasteful feeling that there is still something missing.  Indeed, Quantum
Mechanics provides us with a mathematical framework by which we can derive the observed physics, and
not---as we expect from a theory---a set of physical laws or principles, from which the mathematical
framework is derived. Undeniably the axioms of Quantum Mechanics are of a highly abstract and
mathematical nature, and there is no direct connection between the mathematical formalism and
reality.

\par If one considers the universal validity of Quantum Mechanics, its "physical" axioms---if they
exist---must be of very general nature: they must even transcend Physics itself, moving to the
higher level of Epistemology. Indeed Quantum Mechanics could be regarded itself as a miniature
epistemology, being the {\em quantum measurement} the prototype {\em cognitive act} of interaction
with reality, the epistemic archetype. In this respect the axioms of Quantum Mechanics should be
related to {\em observability principles}, which must be satisfied regardless the specific physical
laws that are object of the experiment. In this search for operational axioms we are also motivated
by the need of understanding the intimate relationships that are logically connecting epistemic issues
such as locality, causality, information-processing complexity, and experimental complexity.  Which
features are really specific to Quantum Mechanics? Or is Quantum Mechanics a logical necessity,
without which we could not even experiment our world?

In a previous work \cite{dariano-losini2005} I showed how it is possible to derive the Hilbert space
formulation of Quantum Mechanics from five operational Postulates concerning {\em experimental
  accessibility and simplicity}. There I showed that the {\em generalized effects} can be
represented as Hermitian matrices over a complex Hilbert space, and I derived a
Gelfand-Naimark-Segal (GNS) representation \cite{GelfandNeumark} for transformations.  The present
paper simplifies and sharpens that derivation, while fixing a subtle error (see Section \ref{s:erra}
on errata). The mathematical formulation of Quantum Mechanics in terms of complex Hilbert space is
derived starting from the five Postulates, for finite dimensions. For the infinite dimensional case
a C${}^*$-algebra representation of physical transformations is derived, starting from just four of
the five Postulates, via a Gelfand-Naimark-Segal (GNS) construction

The starting point for the axiomatization is a seminal definition of {\em physical experiment},
which, as first shown in Ref.  \cite{darianoVax2005}, entails a thorough series of notions that lie
at the basis of the five Postulates. The postulated existence of a {\em faithful state}, which
allows one to calibrate the experimental apparatus, provides operational definitions for the scalar
product and for the {\em transposed} of a transformation.  What is new in the present paper is the
operational deduction of the involution corresponding to the {\em complex-conjugation} for effects,
whose extension to transformations allows to define the usual {\em adjoint} when the extension is
composition-preserving. I will shortly discuss the existence of such composition-preserving
extension among all possible extensions: it is not clear yet if it can be proved in the general
case, or if it will actually require an additional postulate. The operational definition of adjoint
is the core of the derivation of the C${}^*$-algebra representation of physical transformations via
the Gelfand-Naimark-Segal (GNS) construction, which is valid in the generally infinite dimensional
case.

There is a strong affinity of the present work with the program of G.  Ludwig \cite{Ludwig-axI} and
his school (see some papers collected in the book \cite{hartkamper74}). That program didn't succeed
in being an operational axiomatization because it was mainly focused on the convex structure of
quantum theory (which is mathematically quite poor), more than on aspects related to bipartite systems.  In
the present axiomatization some new crucial ingredients---unknown to Ludwig---come from modern Quantum
Tomography \cite{tomo_lecture}, and concern the possibility of performing a complete quantum
calibration of measuring apparatuses \cite{calib} or transformations \cite{tomo_channel} by using a
single pure bipartite state---a so-called {\em faithful state} \cite{faithful}.

\section{The operational axiomatization}
\begin{gaxiom}[On experimental science]\label{ga:1}
  In any experimental science we make {\em experiments} to get {\em information} on the {\em state}
  of a {\em objectified physical system}. Knowledge of such a state will allow us to predict the
  results of forthcoming experiments on the same object system. Since we necessarily work with only
  partial {\em a priori} knowledge of both system and experimental apparatus, the rules for the
  experiment must be given in a probabilistic setting.
\end{gaxiom}
\begin{gaxiom}[On what is an experiment]\label{ga:2} An experiment on an
  object system consists in having it interact with an apparatus. The interaction between object
  and apparatus produces one of a set of possible transformations of the object, each one occurring
  with some probability. Information on the ``state'' of the object system at the beginning of the
  experiment is gained from the knowledge of which transformation occurred, which is the "outcome"
  of the experiment signaled by the apparatus.
\end{gaxiom}
\begin{postulate}[Independent systems]\label{p:independent} There exist independent physical systems.
\end{postulate}
\begin{postulate}[Informationally complete observable]\label{p:infocom} For each physical system
  there exists an informationally complete observable. 
\end{postulate}
\index{local observability principle} 
\begin{postulate}[Local observability principle]\label{p:locobs} For every composite system there exist
  informationally complete observables made only of local informationally complete observables.
\end{postulate}
\begin{postulate}[Informationally complete discriminating observable]\label{p:Bell} For every system
  there exists a minimal informationally complete observable that can be achieved using a joint
  discriminating observable on the system + an ancilla (i.e. an identical independent system). 
\end{postulate}
\begin{postulate}[Symmetric faithful state]\label{p:faith} For every composite system made of two identical
  physical systems there exist a symmetric joint state that is both dynamically and preparationally faithful. 
\end{postulate}
\medskip The General Axioms \ref{ga:1} and \ref{ga:2} entail a very rich series of notions,
including those used in the Postulates---\eg independent systems, observable, informationally
complete observable, etc.  In Sections \ref{s:states} to \ref{s:faithful}, starting from the two
General Axioms, I will introduce step by step such notions, giving the pertaining definitions and
the logically related rules. For a discussion on the General Axioms the reader is addressed to the
publication \cite{darianoVax2005}, where also the generality of the definition of the experiment given
in the General Axiom \ref{ga:1} is analyzed in some detail.
\section{Transformations, States, Independent systems}\label{s:states}
\par Performing a different experiment on the same object obviously
corresponds to use a different experimental apparatus or, at least, to change some apparatus
settings. Abstractly this corresponds to change the set $\{\tA_j\}$ of possible transformations,
$\tA_j$, that the system can undergo.
\glossary{\Idx{transformations1}$\tA,\tB,\ldots,\tA_j,\tB_j,\ldots$ & transformations}
\index{transformation} Such change in practice could mean to alter the "dynamics" of the
transformations, but it may simply mean changing only their probabilities, or, just their labeling.
Any such change actually corresponds to a modification of the experimental setup. Therefore, the set
of all possible transformations $\{\tA_j\}$ will be identified with the choice of experimental
setting, \ie with the {\em experiment} itself---which can be equivalently regarded as the {\em
  "action"} of the experimenter. This will be formalized by the following definition.
\begin{definition}[Experiment]\glossary{\Idx{experiments}$\AA,\AB,\AC,\ldots$ & experiments}
  \index{experiment!definition}\index{outcome} An {\bf experiment} on the object system is
  identified with the set $\AA\equiv\{\tA_j\}$ of possible transformations $\tA_j$ having overall unit
  probability, the apparatus signaling the {\bf outcome} $j$ labeling which transformation
  actually occurred.
\end{definition}
Thus the experiment is just a {\em complete} set of possible transformations that can occur in an
experiment.  In a general cause-and-effect probabilistic framework one shoud regard the experiment
$\AA$ as the "cause" and the {\em outcome} $j$---or the corresponding transformations \index{cause
  and effect} $\tA_j$---as the "effect".\footnote{The reader should not confuse this common usage of
  the word ``effect'' with the homonymous notion used in Sect. \ref{s:effects}.}  The experiment has
to be regarded as the ``cause''---\ie the "action" of the experimenter---since he generally has no
control on which transformation actually occurs, but can decide which experiment to perform, namely
he can choose the set of possible transformations $\AA=\{\tA_j\}$. For example, in an Alice\&Bob
communication scenario Alice will encode different characters by changing the set $\AA$.  The
experimenter has control on the transformation itself only in the special case when the
transformation $\tA$ is deterministic, corresponding to the {\em singleton experiment}
$\AA\equiv\{\tA\}$.  \index{transformation!deterministic} \medskip
\par In the following, wherever we consider a nondeterministic transformation $\tA$ by itself, we
always regard it in the context of an experiment, namely assuming that there always exists at least
a complementary transformation $\tB$ such that the overall probability of $\tA$ and $\tB$ is unit.
Now, according to the General Axiom \ref{ga:1} by definition the knowledge of the state of a physical system
allows us to predict the results of forthcoming possible experiments on the system---more
generally, on another system in the same physical situation.  Then, according to the General Axiom
\ref{ga:2} a precise knowledge of the state of a system would allow us to evaluate the probabilities
of any possible transformation for any possible experiment.  It follows that the only possible
definition of state is the following
\begin{definition}[States]\label{istate}\index{state(s)}
\glossary{\Idx{state1}$\omega,\zeta,\ldots$ & states}
\glossary{\Idx{state2}$\Omega,\Phi,\ldots$ & multipartite states}
 A  state $\omega$ for a physical
  system is a rule that provides the probability for any possible
  transformation, namely 
\begin{equation}
\omega:\textbf{state},\quad\omega(\tA):\text{probability that the
  transformation $\tA$ occurs}.
\end{equation}
In the following for a given physical system we will denote by $\Stset$ the set of all possible
states and by $\Trnset$ the set of all possible transformations.
\glossary{\Idx{convex1}$\Stset$ & convex set of states}
\glossary{\Idx{convex1}$\Trnset$ & truncated convex cone of physical transformations}
\end{definition}
\medskip
We assume that the identical transformation $\tI$ occurs with probability one, namely 
\glossary{\Idx{transformations2}$\tI$ & identical transformation}
\begin{equation}
\omega(\tI)=1.\label{normcond}
\end{equation}
This corresponds to an {\em interaction picture a la Dirac}, in which the free evolution is trivial,
\index{interaction picture} corresponding to a special choice of the lab reference frame (the
scheme, however, could be easily generalized to include a free evolution). Therefore, mathematically
a state will be a map $\omega$ from the set of physical transformations to the interval $[0,1]$,
with Eq. (\ref{normcond}) as a normalization condition. Moreover, for every experiment
$\AA=\{\tA_j\}$ one will have the completeness condition \index{action!normalization condition}
\begin{equation}
\sum_{\tA_j\in\AA}\omega(\tA_j)=1
\end{equation}
for all states $\omega\in\Stset$ of the system. As already noticed in Ref. \cite{darianoVax2005}, in
order to include also non-disturbing experiments, we must conceive situations in which all states
are left invariant by each transformation.  \medskip
\par The fact that we necessarily work in the presence of partial knowledge about both object and
apparatus corresponds to the possibility of a not completely determined specification of both states
and transformations, entailing the convex structure on states and the addition rule for coexistent
transformations.  The addition rule for coexistent transformations will be introduced in Rule
\ref{g:addtrans} in Section \ref{s:transequivalence}. The convex structure of states is given by the
following rule
\begin{grule}[Convex structure of states]\label{idim}\index{state(s)!convex structure}
  The set of possible states $\Stset$ of a physical system is a convex set: for any two states
  $\omega_1$ and $\omega_2$ we can consider the state $\omega$ which is the {\em mixture} of
  $\omega_1$ and $\omega_2$, corresponding to have $\omega_1$ with probability $\lambda$ and
  $\omega_2$ with probability $1-\lambda$. We will write
\begin{equation}
\omega=\lambda\omega_1+(1-\lambda)\omega_2,\quad 0\leq\lambda\leq 1,
\end{equation}
and the state $\omega$ will correspond to the following probability
rule for transformations $\tA$
\begin{equation}
\omega(\tA)=\lambda\omega_1(\tA)+(1-\lambda)\omega_2(\tA).
\end{equation}
\end{grule}
Generalization to more than two states is obtained by induction.
We will call {\em pure} the states which are the extremal elements of
the convex set, namely which cannot be obtained as mixture of any two
states, and we will call {\em mixed} the non-extremal ones. As regards
transformations, the addition of coexistent transformations and
the convex structure will be considered in Rules \ref{g:addtrans} and \ref{r:convextrans}.
\bigskip
\begin{grule}[Transformations form a monoid]\label{isemigroup}
\index{transformation!semigroup of}\index{transformation!composition}
\index{transformation!monoid of}
The composition $\tA\circ\tB$ of two transformations $\tA$ and $\tB$
is itself a transformation. Consistency of composition of transformations requires {\em
associativity}, namely\index{transformation!associativity} 
\begin{equation}
\tC\circ(\tB\circ\tA)=(\tC\circ\tB)\circ\tA.
\end{equation}
There exists the identical transformation $\tI$ which leaves the physical system invariant, and
which for every transformation $\tA$ satisfies the composition rule 
\begin{equation}
\tI\circ\tA=\tA\circ\tI=\tA.
\end{equation}
Therefore, transformations make a semigroup with identity, \ie a {\em monoid}.
\end{grule}
\begin{definition}[Independent systems and local experiments]\label{iindep}
  \index{independent systems}\index{independence}\index{experiment!local} We say that two physical
  systems are {\em independent} if on each system we can perform {\em local experiments}, \ie
  experiments whose transformations commute each other. More precisely, for each transformation
  $\tA^{(1)}\in\AA^{(1)}$ of the local experiment $\AA^{(1)}$ on system 1 and for each transformation
  $\tB^{(2)}\in\AB^{(2)}$ of the local experiment on $\AB^{(2)}$ system 2 one has
\begin{equation}
\tA^{(1)}\circ\tB^{(2)}=\tB^{(2)}\circ\tA^{(1)}.
\end{equation}
\end{definition}
Notice that the above definition of independent systems is purely dynamical, \ie it does not contain
any statistical requirement, such as the existence of factorized states. The present notion of dynamical
independence is so minimal that it can be satisfied not only by the quantum tensor product, but also
by the quantum direct sum. As we will see in the following, it is the local observability principle
of Postulate \ref{p:locobs} which will select the tensor product. It is also worth noticing that in
this operational context appropriate definitions of direct sum and product could be given in a
category theory framework. 

In the following, when dealing with more than one independent system, we will denote local
transformations as ordered strings of transformations as follows
\begin{equation}\label{notlocal}
\tA,\tB,\tC,\ldots\doteq \tA^{(1)}\circ\tB^{(2)}\circ\tC^{(3)}\circ\ldots
\end{equation}
\glossary{\Idx{transformations3a}$(\tA,\tB,\tC,\ldots)$ & local transformations}
\glossary{\Idx{transformations3b}$\tA^{(1)}\circ\tB^{(2)}\circ\tC^{(3)}\circ\ldots$ & local transformations}
\section{Conditioned states and local states}
\begin{grule}[Bayes] When composing two transformations $\tA$ and $\tB$, the probability
$p(\tB|\tA)$ that $\tB$ occurs conditional on the previous occurrence of $\tA$ is given by the Bayes
rule\index{Bayes rule} 
\begin{equation}
p(\tB|\tA)=\frac{\omega(\tB\circ\tA)}{\omega(\tA)}.
\end{equation}
\end{grule}
The Bayes rule leads to the concept of {\em conditional
  state}:\index{state(s)!conditional}\index{conditional state}
\begin{definition}[Conditional state]\label{istatecond} The {\em conditional
state} $\omega_\tA$ gives the probability that a transformation
$\tB$ occurs on the physical system in the state $\omega$ after the
transformation $\tA$ has occurred, namely
\begin{equation}\label{condstate}
\omega_\tA(\tB)\doteq\frac{\omega(\tB\circ\tA)}{\omega(\tA)}.
\end{equation}
\glossary{\Idx{state3}$\omega_\tA$ & conditional state (state $\omega$ conditioned by the
  transformation $\tA$)}
\end{definition}
\par In the following we will make extensive use of the functional notation
\begin{equation}
\omega_\tA\doteq\frac{\omega(\cdot\circ\tA)}{\omega(\tA)},
\end{equation}
where the centered dot stands for the argument of the map. Therefore, the notion of conditional state describes
the most general {\em evolution}. 
\begin{definition}[Local state]\label{istateloc}
  \index{state(s)!local}\index{local!state} In the presence of many independent systems in a joint
  state $\Omega$, we define the {\bf local state} $\Omega|_n$ of the $n$-th system as the
  probability rule of the joint state $\Omega$ with a local transformation $\tA$ only on the $n$-th
  system and with all other systems untouched, namely
\begin{equation}
\Omega|_n(\tA)\doteq\Omega(\tI,\ldots,\tI,\underbrace{\tA}_{n\text{th}},\tI,\ldots).
\end{equation}
\glossary{\Idx{state4}$\Omega|_n$ & local state}
\end{definition}
For example, for two systems only, (or, equivalently, grouping $n-1$ systems into a single one), we
just write $\Omega|_1=\Omega(\cdot,\tI)$.
\begin{remark}[Linearity of evolution]
  The present definition of ``state'', which logically follows from the definition of experiment,
  leads to the identification {\em state-evolution}$\equiv${\em state-conditioning}, entailing a
  {\em linear action of transformations on states}, apart from normalization. In addition, since
  states are probability functionals on transformations, by dualism (equivalence classes of)
  transformations will be identified as linear functionals over the state space.
\end{remark}
\bigskip
\glossary{\Idx{state2}$\tilde\omega,\tilde\zeta,\ldots$ & weights}
\par It is convenient to extend the notion of state to that of {\em weight}, \index{weight} 
\ie a nonnegative bounded functionals $\tilde\omega$ over the set of transformations with
$0\leq\tilde\omega(\tA)\leq\tilde\omega(\tI)<+\infty$ for all transformations $\tA$.  To each weight
$\tilde\omega$ it corresponds the properly normalized state
\begin{equation}
\omega=\frac{\tilde\omega}{\tilde\omega(\tI)}.
\end{equation}

Weights make the convex cone $\Wset$ generated by the convex set of states $\Stset$.
\glossary{\Idx{convex}$\Wset$ & convex cone $\Wset$ generated by the convex set of states}
\begin{definition}[Linear real space of generalized weights]
  \index{generalized weights} We extend the notion of weight to that of negative weight, by taking
  differences. Such generalized weights span the affine linear space $\Wset_\Reals$ of the convex
  cone $\Wset$ of weights.
\end{definition}
\begin{remark} The transformations $\tA$ act as linear transformations over the space of weights as
  follows
\begin{equation}
\tA \tilde\omega=\tilde\omega(\tB\circ\tA).\label{wtransf} 
\end{equation}
\end{remark}

We are now in position to introduce the concept of {\em operation}.
\begin{definition}[Operation]\label{operation}\index{operation} 
To each transformation $\tA$ we can associate a linear map $\Op{\tA}:\;\Stset\longrightarrow\Wset$,
which sends a state $\omega$ into the unnormalized state $\tilde\omega_\tA\doteq
\Op{\tA}\omega\in\Wset$, with $\tilde\omega_\tA(\tB)=\omega(\tB\circ\tA)$, namely
\begin{equation}\label{Schpict}
\tA\omega:=\omega(\cdot\circ\tA)\equiv\Op{\tA}\omega\equiv\tilde\omega_\tA.
\end{equation}
\end{definition}
This is the analogous of the Schr\"{o}dinger picture evolution of states in Quantum Mechanics. One
can see that in the present context linearity of evolution is just a consequence of the fact that
the evolution of states is pure state-conditioning: this will includes also the deterministic case
$\tU\omega=\omega(\cdot\circ\tU)$ of transformations $\tU$ with $\omega(\tU)=1$ for all states
$\omega$---the analogous of unitary evolutions and channels in Quantum Mechanics. More generally,
the operation $\operatorname{Op}$ gives both the conditioned state and the probability of the
transformation as follows
\index{state!state-reduction} \index{state-reduction}
\begin{equation}
\omega_\tA\equiv\frac{\Op{\tA}\omega}{\Op{\tA}\omega(\tI)},\qquad
\omega(\tA)\equiv\Op{\tA}\omega(\tI).
\end{equation}
\glossary{\Idx{operation}$\Op{\tA}$ & operation corresponding to the transformation $\tA$}
\bigskip
\index{transformation!local}\index{local!transformation}
\section{Dynamical and informational structure}\label{s:transequivalence}
From the Bayes rule, or, equivalently, from the definition of
conditional state, we see that we can have the following complementary situations:
\begin{enumerate}
\item There are different transformations which produce the same state
  change, but generally occur with different probabilities;
\item There are different transformations which always occur with the
  same probability, but generally affect a different state change.
\end{enumerate}
The above observation leads us to the following definitions of
dynamical and informational equivalences of transformations.
\begin{definition}[Dynamical equivalence of transformations]\label{d:dyneq}
  \index{transformation!dynamical equivalence} \index{dynamical equivalence of transformations} Two
  transformations $\tA$ and $\tB$ are dynamically equivalent if $\omega_\tA=\omega_\tB$ for all
  possible states $\omega$ of the system.
\end{definition}
  We will denote the equivalence class containing the  transformation $\tA$ as $[\tA]_\dyn$.
\medskip
\begin{definition}[Informational equivalence of transformations]
  \index{transformation!informational equivalence} \index{informational equivalence of
    transformations} Two transformations $\tA$ and $\tB$ are informationally equivalent if
  $\omega(\tA)=\omega(\tB)$ for all possible states $\omega$ of the system.
\end{definition}
  We will denote the equivalence class containing the transformation $\tA$ as $[\tA]_\eff$, since,
  as we will see in the following, such equivalence class will be identified with the notion of {\em
    effect}. 
\begin{definition}[Identification of transformations/experiments]\label{d:compleq}
  \index{transformation!complete equivalence} \index{experiment!complete equivalence}
  \index{complete equivalence!of transformations} \index{complete equivalence!of experiments} Two
  transformations (or experiments) are completely equivalent iff they are both dynamically and
  informationally equivalent, and we will simply say that the two transformations are equal.
\end{definition}
\begin{theorem}[Identity of transformations]\label{t:eqtrans} Two transformations $\tA_1$ and
  $\tA_2$ are identical if and only if one has
\begin{equation}\label{idtransf}
\omega(\tB\circ\tA_1)=\omega(\tB\circ\tA_2), \;\forall\omega\in\Stset,\,\forall\tB\in\Trnset.
\end{equation}
\end{theorem}
\Proof Identity (\ref{idtransf}) for $\tB=\tI$ is the informational equivalence of $\tA_1$ and
$\tA_2$. On the other hand, since $\omega(\tA_1)=\omega(\tA_2)$ $\forall\omega\in\Stset$,
Eq. (\ref{idtransf}) also implies that 
\begin{equation}
\omega_{\tA_1}=\omega_{\tA_2},\;\forall\omega\in\Stset,
\end{equation}
namely the two transformations are also dynamically equivalent, whence they are completely
equivalent.\qed Notice that even though two transformations are completely equivalent, in principle
they can still be experimentally different, in the sense that they are achieved with different
apparatus. However, we emphasize that outcomes in different experiments corresponding to completely
equivalent transformations always provide the same information on the state of the object, and,
always produce the same conditioning of the state.  

\par The notions of dynamical and informational equivalences of transformations leads one to
introduce a convex structure also for transformations.  We first need the notion of informational
compatibility.  \index{transformation!coexistence} \index{transformation!informational
  compatibility} \index{informational compatibility of transformations} \index{coexistence of
  transformations}
\begin{definition}[Informational compatibility or coexistence] We say that
two transformations $\tA$ and $\tB$ are {\em coexistent} or {\em
informationally compatible} if one has 
\begin{equation}
\omega(\tA)+\omega(\tB)\leq 1,\quad\forall\omega\in\Stset.\label{compatible}
\end{equation}
\end{definition}
The fact that two transformations are coexistent means that, in principle, they can occur in the
same experiment, namely there exists at least an experiment containing both of them. We have named
the present kind of compatibility "informational" since it is actually defined on the informational
equivalence classes of transformations.
\par We are now in position to define the "addition" of coexistent transformations.
\begin{grule}[Addition of coexistent transformations]\label{g:addtrans}
\index{transformation!addition} For any two
  coexistent transformations $\tA$ and $\tB$  we define the
  transformation $\tS=\tA_1+\tA_2$ as the transformation corresponding
  to the event $e=\{1,2\}$, namely the apparatus signals that either
  $\tA_1$ or $\tA_2$ occurred, but does not specify which one.
By definition, one has
\begin{equation}\label{r:sum1}
\forall\omega\in\Stset\qquad\omega(\tA_1+\tA_2)=\omega(\tA_1)+\omega(\tA_2),
\end{equation}
whereas the state conditioning is given by
\begin{equation}\label{r:sum2}
\forall\omega\in\Stset\qquad
\omega_{\tA_1+\tA_2}=\frac{\omega(\tA_1)}{\omega(\tA_1+\tA_2)}
\omega_{\tA_1}+\frac{\omega(\tA_2)}{{\omega(\tA_1+\tA_2)}}\omega_{\tA_2}.
\end{equation}
\glossary{\Idx{transformations5}$\tA+\tB$ & addition of compatible transformations}
\end{grule}
Notice that the two rules in Eqs.  (\ref{r:sum1}) and (\ref{r:sum2}) completely specify the 
transformation $\tA_1+\tA_2$, both informationally and dynamically. Eq. (\ref{r:sum2}) can be more
easily restated in terms of operations as follows:
\begin{equation}\label{r:addtrans}
\forall\omega\in\Stset\qquad
(\tA_1+\tA_2)\omega=\tA_1\omega+\tA_2\omega.
\end{equation}
It is easy to check that the composition "$\circ$" of transformations is distributive with respect
to the addition "$+$".  Addition of compatible transformations is the core of the description of
partial knowledge on the experimental apparatus. Notice also that the same notion of coexistence can
be extended to "effects" as well (see Definition \ref{d:effect}). In the following we will use the
notation
\begin{equation}\label{channelA}
\tS(\AA):=\sum_{\tA_j\in\AA} \tA_j
\end{equation}
to denote the deterministic transformation $\tS(\AA)$ that corresponds to the sum of all possible
transformations $\tA_j$ in $\AA$. 

\index{acausality of local actions}
\bigskip
\par At first sight it is not obvious that the commutativity of local transformations in Definition
\ref{iindep} implies that a local "action" on system 2 does not affect the conditioned local state
on system 1. Indeed, the occurrence of the transformation $\tB$ on system 1 generally affects the
local state on system 2, i. e. $\Omega_{\tB,\tI}|_2\neq\Omega_2$. However, local "actions" on a
system have no effect on another independent system, as it is proved in the following theorem.

\begin{theorem}[No signaling, \ie acausality of local actions]\label{iacausal}
  Any local "action" (\ie experiment) on a system does not affect another independent system. More
  precisely, any local action on a system is equivalent to the identity transformation when viewed
  from another independent system. In equations one has
\begin{equation}
\forall\Omega\in\Stset^{\times 2},\forall\AA,\qquad
\Omega_{\tS(\AA),\tI}|_2=\Omega|_2.
\end{equation}
\end{theorem}
\Proof By definition, for $\tB\in\Trnset$ one has $\Omega|_2(\tB)=\Omega(\tI,\tB)$, and using
Eq. (\ref{channelA}) according to Rule \ref{g:addtrans} one has
\begin{equation}
\Omega(\tS(\AA),\tB)=\sum_{\tA_j\in\AA}\Omega(\tA_j,\tB)=\Omega(\tI,\tB)=:\Omega|_2(\tB).
\end{equation}
On the other hand, we have
\begin{equation}
\Omega_{\tS(\AA),\tI}|_2(\tB)=\Omega((\tI,\tB)\circ(\tS(\AA),\tI)=
\Omega(\tS(\AA),\tB),
\end{equation}
namely the statement.\qed
\bigskip
Notice the consistency with Rule \ref{g:addtrans}:
\begin{equation}
\begin{split}
\Omega_{\tS(\AA),\tI}|_2(\tB)=&\Omega_{\tS(\AA),\tI}(\tI,\tB)=
\sum_{\tA_j\in\AA}\Omega_{\tA_j,\tI}(\tI,\tB)\frac{\Omega(\tA_j,\tI)}{
\sum_{\tA_j\in\AA}\Omega(\tA_j,\tI)}\\
=&\sum_{\tA_j\in\AA}\frac{\Omega(\tA_j,\tB)}{\Omega(\tA_j,\tI)}\frac{\Omega(\tA_j,\tI)}{
\Omega(\tI,\tI)}=\sum_{\tA_j\in\AA}\Omega(\tA_j,\tB)=\Omega(\tI,\tB).
\end{split}
\end{equation}

It is worth noticing that the no-signaling is a mere consequence of our minimal notion of dynamical
independence in Def. \ref{iindep}.

\begin{grule}[Multiplication of a transformation by a scalar]\label{g:scalmult}
\index{transformation!multiplication by a scalar}
For each transformation $\tA$ the transformation $\lambda\tA$ for
$0\leq\lambda\leq 1$ is defined as the transformation which is 
dynamically equivalent to $\tA$, but which occurs with probability
$\omega(\lambda\tA)=\lambda\omega(\tA)$.   
\end{grule}
Notice that according to Definition \ref{d:compleq} two transformations are completely characterized
operationally by the informational and dynamical equivalence classes to which they belong, whence
Rule \ref{g:scalmult} is well posed. 

\bigskip Clearly $\lambda\tA_1$  and $(1-\lambda)\tA_2$ are coexistent
$\forall\tA_1,\tA_2\in\Trnset$, $\lambda\in[0,1]$. We can therefore pose a convex structure over the
set of physical transformations $\Trnset$.
\begin{grule}[Convex structure of physical transformations]\label{r:convextrans}
  \index{transformation!convex structure} The set $\Trnset$ of physical transformations is convex,
  namely for any two physical transformations $\tA_1$ and $\tA_2$ we can consider the physical
  transformation $\tA$ which is the {\em mixture} of $\tA_1$ and $\tA_2$ with probabilities
  $\lambda$ and $1-\lambda$.  Formally we write
\begin{equation}
\tA=\lambda\tA_1+(1-\lambda)\tA_2,\quad 0\leq\lambda\leq 1,\label{ala}
\end{equation}
with the following meaning: the physical transformation $\tA$ is itself a probabilistic
transformation, occurring with overall probability
\begin{equation}
\omega(\tA)=\lambda\omega(\tA_1)+(1-\lambda)\omega(\tA_2),
\end{equation}
meaning that when the transformation $\tA$ occurred we know that the transformation dynamically was
either $\tA_1$ with (conditioned) probability $\lambda$ or $\tA_2$ with probability $(1-\lambda)$.
\end{grule}
\medskip

As we will see in Section \ref{s:Banach}, the convex set of physical transformations $\Trnset$ has
the form of a truncated convex cone in the Banach algebra of generalized transformations.

\begin{remark}[Algebra of generalized transformations]\label{rem:transf}
  Using Eqs. (\ref{r:sum1}) and (\ref{r:addtrans}) one can extend the addition of coexistent
  transformations to generic linear combinations, that we will call {\em generalized
    transformations} \index{generalized transformations} (to be contrasted with the original notion,
  for which we will keep the name {\em physical transformations}). The generalized transformations
  constitute a real vector space---hereafter denoted as $\Trnset_\Reals$---which is the affine space
  of the convex space $\Trnset$. Composition of transformations can be extended via linearity to
  generalized transformations, making their space a real algebra, the {\em algebra of generalized
    transformations}.
\end{remark}
\begin{remark}[Cone and double-cone of generalized transformations]\label{r:dcone} The generalized 
  transformations $\tG$ of the form $\tG=\lambda\tA$ with $\tA$ physical transformation and
  $\lambda\geq 0$ make a cone (denoted by $\Trnset_\Reals^+$), and for $\lambda\in\Reals$ make a
  double cone (denoted by $\Trnset_\Reals^\pm$). Notice that for
  $\Trnset_\Reals\ni\tG\not\in\Trnset_\Reals^\pm$ \ie out of the double cone the conditioning
  $\omega_\tG$ is not necessarily a state (\eg there exist a physical transformation $\tA$ for which
  $\omega_\tG(\tA)>1$ or $\omega_\tG(\tA)<0$, even though $\omega_\tG(\tI)=1$. On the other
  hand, for generalized transformations in the double cone $\omega_\tG$ is always a true state.
\end{remark}
Indeed, for a generalized transformation $\tG=\lambda\tA\in\Trnset_\Reals^\pm$ proportional to a
physical transformation $\tA$ one has 
\begin{equation}
\omega_\tG(\tB)=\frac{\omega(\tB\circ\tG)}{\omega(\tG)}=
\frac{\omega(\tB\circ\lambda\tG)}{\omega(\lambda\tG)}=\frac{\omega(\tB\circ\tA)}{\omega(\tA)}.
\end{equation}
However, for a generalized transformation $\tG=\tA_1-\tA_2\not\in\Trnset_\Reals^\pm$
with $\cA_1\neq\cA_2$ one has
\begin{equation}
\omega_{\tA_1-\tA_2}=\frac{\omega(\tA_1)}{\omega(\tA_1)-\omega(\tA_2)}\omega_{\tA_1}-
\frac{\omega(\tA_2)}{\omega(\tA_1)-\omega(\tA_2)}\omega_{\tA_2}=
\lambda\omega_{\tA_1}+(1-\lambda)\omega_{\tA_2},
\end{equation}
and, generally one can have $\lambda> 1$, in which case consider \eg a transformation $\tB$ for which
$\omega_{\tA_1}(\tB)\geq \lambda^{-1}$ and $\omega_{\tA_2}(\tB)=0$. Then, one has
$\omega_{\tA_1-\tA_2}(\tB)> 1$.
\section{Effects\label{s:effects}}
\glossary{\Idx{effect1}$\cA,\cB,\ldots,$ & effects}
\glossary{\Idx{effect2}$[\tA]_\eff$ & effect containing the transformation $\tA$}
Informational equivalence leads to the notion of {\em effect}, which corresponds closely to the same
notion introduced by Ludwig \cite{Ludwig-axI}.\footnote{In previous
  literature \cite{darianoVax2005} I adopted the name "{\em propensity}" for the informational
  equivalence class of transformations. The intention was to keep a separate word, since the world 
"effect" has already been identified with the quantum mechanical notion, corresponding to a precise
mathematical object (\ie a positive contraction). 
However, it turned out that the adoption of the world ``propensity'' has the negative effect of
linking the present axiomatic with the Popperian interpretation of probability.}
\begin{definition}[Effects]\label{d:effect}
\index{effect} We call {\bf effect} an  informational equivalence class of transformations.  
\end{definition}

In the following we will denote effects with the underlined symbols $\cA$, $\cB$, etc., and we will
use the same notation to denote the effect containing the transformation $\tA$, \ie $\tA_0\in\cA$
means "$\tA_0$ is informationally equivalent to $\tA$" (depending on convenience we will also keep
the notation $[\tA]_\eff$).  Thus, by definition one has $\omega(\tA)\equiv\omega(\cA)$, and we will
legitimately write $\omega(\cA)$. Similarly, one has $\tilde\omega_\tA(\tB)\equiv
\tilde\omega_\tA(\cB)$ which implies that $\omega(\tB\circ\tA)=\omega(\cB\circ\tA)$ which gives the
chaining rule
\begin{equation}
\cB\circ\tA\subseteq[\tB\circ\tA]_\eff,
\end{equation}
corresponding to the "Heisenberg picture" version of Eq. (\ref{Schpict}), with the operation
$\Op{\tA}$ acting on effects $\cB$, namely
\begin{equation}\label{operatingoneffect}
\Op{\tA} \cB :=\cB\circ\tA.
\end{equation}
One also has the locality rule
\begin{equation}
[(\tA,\tB)]_\eff\supseteq([\tA]_\eff,[\tB]_\eff).
\end{equation}
using notation (\ref{notlocal}).  It is clear that $\lambda\tA$ and $\lambda\tB$ belong to
the same equivalence class iff $\tA$ and $\tB$ are informationally equivalent. This means that also
for effects multiplication by a scalar can be defined as $\lambda\cA=[\lambda\tA]_\eff$.  Moreover,
we can naturally extend the notion of coexistence from transformation to effects, and for
$\tA_0\in\cA$ and $\tB_0\in\cB$ one has $\tA_0+\tB_0\in[\tA+\tB]_\eff$, we can define addition of
coexistent effects as \index{effect!addition} $\cA+\cB=[\tA+\tB]_\eff$ for any choice of
representatives $\tA$ and $\tB$ of the two added effects. We will denote the set of effects by
$\Cntset$. We will also extend the notion of effect to that of {\em generalized effects} by taking
differences of effects \index{generalized effects} (for the original notion, we will use the name
{\em physical effects}).  The set of generalized effects will be denoted as $\Cntset_\Reals$.

\begin{grule}[Convex set of physical effects] In a way completely analogous to Rule
  \ref{r:convextrans} the set of physical effects $\Cntset$ is convex.
\end{grule}

\section{The real Banach space structure}\label{s:Banach}

\begin{theorem}[Banach space $\Cntset_\Reals$ of generalized effects]\label{t:effectnorm} The
  generalized effects make a Banach space, with norm defined as follows \index{effect!norm}
  \glossary{\Idx{effect}$\n{\cA}$ & norm of effects}
\begin{equation}\label{effectnorm}
\n{\cA}=\sup_{\omega\in\Stset}|\omega(\cA)|.
\end{equation}
\index{transformation!contraction}
\end{theorem}
\Proof We remind the axioms of norm:
i) Sub-additivity $\n{\cA+\cB}\leq\n{\cA}+\n{\cB}$; ii) Multiplication by scalar
$\n{\lambda\cA}=|\lambda|\n{\cA}$; iii) $\n{\cA}=0$ implies $\cA=0$.
The quantity in Eq. (\ref{effectnorm}) satisfy the sub-additivity relation i), since
\begin{equation}
\n{\cA+\cB}=\sup_{\omega\in\Stset}|\omega(\cA)+\omega(\cB)|\le
\sup_{\omega\in\Stset}|\omega(\cA)+\sup_{\omega'\in\Stset}|\omega'(\cB)|=
\n{\cA}+\n{\cB}.
\end{equation}
Moreover, it obviously satisfies axiom ii). Finally, axiom iii) corresponds to a generalized effect
that is the (multiple of a) difference of two informationally equivalent transformations, namely the
null effect. Closure with respect to the norm (\ref{effectnorm}) makes the real vector space
$\Cntset_\Reals$ of generalized effects a Banach space, which we will name the {\em Banach space of
  generalized effects}.  The norm closure corresponds to assume preparability of effects by an
approximation criterion in-probability (see also Remark \ref{r:closure}).\qed \medskip
\begin{theorem}[Banach space $\Wset_\Reals$ of generalized weights] The generalized weights make a
  Banach space, with norm defined as follows
\begin{equation}
\n{\tilde\omega}:=\sup_{\cA\in\Cntset_\Reals,\n{\cA}\leq 1}|\tilde\omega(\cA)|.\label{normstate}
\end{equation}
\end{theorem}
\Proof The quantity in Eq. (\ref{normstate}) satisfies the sub-additivity relation 
$\n{\tilde\omega+\tilde\zeta}\leq\n{\tilde\omega}+\n{\tilde\zeta}$, since 
\begin{equation}
\begin{split}
\n{\tilde\omega+\tilde\zeta}=&\sup_{\cA\in\Cntset_\Reals,\n{\cA}\leq 1}|\tilde\omega(\cA)+\zeta(\cA)|
\leq\sup_{\cA\in\Cntset_\Reals,\n{\cA}\leq 1}[|\tilde\omega(\cA)|+|\tilde\zeta(\cA)|]\\ \le&
\sup_{\cA\in\Cntset_\Reals,\n{\cA}\leq
  1}|\tilde\omega(\cA)|+\sup_{\cA\in\Cntset_\Reals,\n{\cA}\leq
  1}|\tilde\zeta(\cA)]|=\n{\tilde\omega}+\n{\tilde\zeta}. 
\end{split}
\end{equation}
Moreover, it obviously satisfies the identity
\begin{equation}
\n{\lambda\tilde\omega}=|\lambda|\n{\omega}.
\end{equation}
Finally, $\n{\tilde\omega}=0$ implies that $\tilde\omega=0$, since either $\tilde\omega$ is a
positive linear form, \ie it is proportional to a true state, whence at least $\tilde\omega(\tI)>0$,
or $\tilde\omega$ is the difference of two positive linear forms, whence the two corresponding
states must be equal by definition, since their probability rules are equal, which means that,
again, $\tilde\omega=0$.
  Closure with respect to the norm (\ref{normstate}) makes the real vector space of generalized
  weights $\Wset_\Reals$ a Banach space, which we will name the {\em Banach space of generalized weights}.
  The norm closure corresponds to assume preparability of states by an approximation
  criterion in-probability (see also Remark \ref{r:closure}).\qed
\medskip
\begin{remark}[Duality between the convex sets of states and of effects]\label{r:duality}
  From the Definition \ref{istate} of state it follows that the convex set of states $\Stset$ and
  the convex sets of effects $\Cntset$ are dual each other, and the latter can be regarded as the
  truncated convex cone of positive linear contractions over the set of states, namely the set of
  bounded positive functionals $l\leq 1$ on $\Stset$, and with the functional $l_{\cA}$
  corresponding to the effect $\cA$ defined as follows \glossary{\Idx{effect4}$l_{\cA}$ & effect
    containing the transformation $\tA$}
\begin{equation}
l_{\cA}(\omega)\doteq\omega(\cA).
\end{equation}
The above duality naturally extends to generalized effects and generalized weights.  Therefore,
$\Wset_\Reals$  and $\Cntset_\Reals$ are a dual Banach pair.
\end{remark}
The above duality is the analogous of the duality between bounded operators and trace-class operators in
Quantum Mechanics. It is worth noticing that this dual Banach pair is just a consequence of the
probabilistic structure that is inherent in our definition of experiment.

In the following we will often identify generalized effects with their corresponding functionals,
and denote them by lowercase letters $a,b,c,\ldots$, or $l_1,l_2,\ldots$ 

\medskip\par  For generalized transformations, a suitably defined norm is the following.
\begin{theorem}[Banach algebra $\Trnset_\Reals$ of generalized transformations]\label{t:transnorm} 
  The set of generalized transformations make a Banach algebra, with norm defined as follows
  \index{transformation!norm} \glossary{\Idx{transformations4}$\n{\tA}$ & norm of transformation}
\begin{equation}\label{wernernorm}
\n{\tA}:=\sup_{\cB\in\Cntset_\Reals,\n{\cB}\leq 1}\n{\cB\circ\tA}
\equiv\sup_{\Cntset_\Reals\ni\n{\cB}\leq 1}\;\sup_{\omega\in\Stset}\;
|\omega(\cB\circ\tA)|.
\end{equation}
\index{transformation!contraction}
\end{theorem}
\Proof For $x\in{\mathcal B}$ in a generic Banach space ${\mathcal B}$ and $T$ a map on ${\mathcal
  B}$ one has $\n{Tx}\leq \n{T}\n{x}$, with $\n{T}:=\sup_{\n{y}\leq 1}\n{Ty}$, and applying the
bound twice one has that for $A$ and $B$ maps on ${\mathcal B}$ one has $\n{AB}\leq\n{A}\n{B}$. In
our case this bound will rewrite $\n{\tB\circ\tA}\leq\n{\tB}\n{\tA}$, whence the generalized
transformations make a Banach algebra. 

It is also clear that, by definition, for each physical
transformation $\tA$ one has $\n{\tA}\leq1$, namely physical transformations are contractions.  The
norm closure corresponds to assume preparability of transformations by an approximation criterion
in-probability (see also Remark \ref{r:closure}).\qed

\begin{theorem}[Bound between the norm of a transformation and the norm of its effect] The following
  bound holds 
\begin{equation}\label{bdnorms}
\n{\cA}\leq\n{\tA}.
\end{equation}
and for transformation $\tA\in\Trnset_\Reals^\pm$ one has the identity
\begin{equation}\label{idnorms}
\n{\cA}=\n{\tA}.
\end{equation}
\end{theorem}
\Proof One can easily check the bound
\begin{equation}
\n{\cA}=\sup_{\omega\in\Stset}|\omega(\tA)|\leq\sup_{\omega\in\Stset,\cC\in\Cntset_\Reals,
\;\n{\cC}\leq 1}|\omega(\cC\circ\tA)|=\n{\tA}.
\end{equation}
For $\tA\in\Trnset_\Reals^\pm$ the generalized weight $\omega_\tA$ is a physical state, and also
the reverse bound holds
\begin{equation}
\begin{split}
\n{\tA}=&\sup_{\omega\in\Stset,\cC\in\Cntset,\n{\cC}\leq1}
|\omega(\cC\circ\tA)|=\sup_{\omega\in\Stset,\cC\in\Cntset,\n{\cC}\leq1}|\omega_\tA(\cC)
\omega(\tA)|\\
\leq&\sup_{\omega\in\Stset}|\omega(\tA)|=\sup_{\omega\in\Stset}|\omega(\cA)|=\n{\cA}, 
\end{split}
\end{equation}
which then implies identity (\ref{idnorms}).
\qed 
\begin{corollary}\label{c:contraction} Two physical transformations $\tA$ and $\tB$ are
  coexistent iff $\tA+\tB$ is a contraction.
\end{corollary}
\Proof If the two transformations are coexistent, then from Eqs. (\ref{compatible}) and
(\ref{wernernorm}) one has that $\n{\tA+\tB}\leq 1$. On the other hand, if $\n{\tA+\tB}\leq 1$, this
means that for all states one has $\omega(\tA)+\omega(\tB)\leq 1$, namely the transformations are
coexistent.\qed
\begin{corollary}Physical transformations are contractions, namely they make a truncated convex
  cone. 
\end{corollary}
\Proof It is an immediate consequence of Corollary \ref{c:contraction}.\qed
\glossary{\Idx{convex5}$\Cntset$ & truncated convex cone of effects}
\glossary{\Idx{effect3}$l$ & effect}
\begin{remark}[Approximability criteria and norm closure]\label{r:closure} The above defined norms
  operationally correspond to approximability criteria in-probability.  The norm closure may not be
  required operationally, however, as any other kind of extension, it is mathematically convenient.
\end{remark}
\bigskip
\section{Observables}
\begin{definition}[Observable]\index{observable} We call observable a complete set of effects
  $\AL=\{l_i\}$ of an experiment $\AA=\{\tA_j\}$, namely one has $l_i=\underline{\tA_j}$ $\forall j$.
\end{definition}
Clearly, the generalized observable is normalized to the constant unit functional, \ie $\sum_il_i=1$.
\begin{definition}[Informationally complete observable] An observable
  $\AL=\{l_i\}$ is informationally complete if each effect can be written as a linear combination of
  the of elements of $\AL$, namely for each effect $l$ there exist coefficients $c_i(l)$ such that
\begin{equation}
l=\sum_ic_i(l)l_i.
\end{equation}
We call the informationally complete observable {\em minimal} \index{informationally
  complete observable!minimal} when its effects are linearly independent.
\end{definition}
\begin{remark}[Bloch representation]\label{r:Bloch} Using an informationally complete observable one can
  reconstruct any state $\omega$ from just the probabilities $l_i(\omega)$, since one has
\begin{equation}
\omega(\cA)=\sum_ic_i(l_{\underline{\tA}})l_i(\omega).
\end{equation}
\end{remark}
\begin{definition}[Predictability and resolution]\label{def:res} 
\index{transformation!predictable}\index{predictable!transformation}
\index{effect!predictable}\index{predictable!effect}
We will call a transformation $\tA$---and likewise its
effect---{\em predictable} if there exists a state for which
$\tA$ occurs with certainty and some other state for which it never
occurs. The transformation (effect) will be also called {\em
resolved} if the state for which it occurs with certainty is
unique---whence pure.
An experiment will be called {\em predictable} when it is made only
of predictable transformations, and {\em resolved} when all
transformations are resolved.
\end{definition}
The present notion of predictability for effects corresponds to that of
"decision effects" of Ludwig \cite{Ludwig-axI}. For a predictable
transformation $\tA$ one has $\n{\tA}=1$. Notice that 
a predictable transformation is not deterministic, and it can
generally occur with nonunit probability on some state $\omega$. 
Predictable effects $\tA$ correspond to affine functions
$f_\tA$ on the state space $\Stset$ with $0\leq f_\tA\leq 1$ achieving
both bounds.

\begin{definition}[Perfectly discriminable set of states] We call a set of states $\{\omega_n\}_{n=1,N}$
  {\em perfectly discriminable} if there exists an experiment $\AA=\{\tA_j\}_{j=1,N}$ with transformations
  corresponding to predictable effects $\cA_j$ satisfying the relation \index{state(s)!perfectly discriminable}
\begin{equation}
\omega_m(\cA_n)=\delta_{nm}.
\end{equation}
\end{definition}
\begin{definition}[Informational dimensionality] We call {\em informational dimension}
\index{dimension!informational} of the convex set of states $\Stset$, denoted by $\idim{\Stset}$,
the maximal cardinality of perfectly discriminable set of states in $\Stset$.
\glossary{\Idx{dimension3}$\idim{\Stset}$ & informational dimension of the convex set of states $\Stset$}
\end{definition}
\begin{definition}[Discriminating observable]
  \index{observable!discriminating}\index{effect!observable} An observable $\AL=\{l_j\}$ is {\em
    discriminating for} $\Stset$ when $|\AL|\equiv\idim{\Stset}$, \ie $\AL$ discriminates a maximal
  set of discriminable states.
\end{definition}
\bigskip
\section{Faithful state}\label{s:faithful}
\begin{definition}[Dynamically faithful state]\label{def:faith} 
\index{state!dynamically faithful}\index{dynamically faithful state}
We say that a state $\Phi$ of a composite system is {\em dynamically faithful} for the $n$th component
system when for every transformation $\tA$ the following map is one-to-one  
\glossary{\Idx{dyn}$[\tA]_\dyn$ & dynamical equivalence class of transformations $\tA$}
\begin{equation}
\tA\leftrightarrow(\tI,\ldots,\tI,\underbrace{\tA}_{n\text{th}},\tI,\ldots)\Phi,
\end{equation}
where in the above equation the transformation $\tA$ acts locally only on the $n$th component system.
\end{definition}
Notice that by linearity the correspondence is still one-to-one when extended to generalized
transformations. Physically, the definition corresponds to say that the output conditioned state
(multiplied by the probability of occurrence) is in one-to-one correspondence with the transformation.
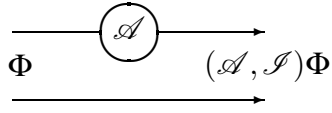
\begin{figure}[hbt]
    \setlength{\unitlength}{800sp}
    \begin{picture}(8745,3219)(931,-3565)
      {\thicklines \put(5401,-1261){\oval(1756,1756)}}
      {\put(1801,-1261){\line(1, 0){2700}}}
      {\put(6301,-1261){\vector(1, 0){3300}}}
      {\put(1801,-3361){\vector(1, 0){7800}}}
      \put(2026,-2611){\makebox(0,0)[b]{$\Phi$}}
      \put(5401,-1486){\makebox(0,0)[b]{$\tA$}}
      \put(9676,-2811){\makebox(0,0)[b]{$(\tA,\tI)\Phi$}}
    \end{picture}
\caption{Illustration of the notion of dynamically faithful state for a bipartite system (see
  Definition \ref{def:faith}). Physically, the state $\Phi$ is faithful when the output conditioned state
(multiplied by the probability of occurrence) is in one-to-one correspondence with the transformation.}
  \end{figure}
\medskip
\par In the following we restrict attention to bipartite systems. In equations a state is dynamically
faithful (for system 1) when
\begin{equation}
(\tA,\tI)\Phi=0\;\Longleftrightarrow\tA=0,
\end{equation}
and according to Definition \ref{operation} this is equivalent to say that for every bipartite
effect $\cB$ one has
\begin{equation}
\Phi(\cB\circ(\tA,\tI))=0\quad\Longleftrightarrow\quad\tA=0.
\end{equation}
\begin{definition}[Preparationally faithful state]\label{prepfaith}
  \index{state!preparationally faithful}\index{preparationally faithful state} We will call a state
  $\Phi$ of a bipartite system {\em preparationally faithful} for system 1 if every joint bipartite
  state $\Omega$ can be achieved by a suitable local transformation $\tT_\Omega$ on system 1
  occurring with nonzero probability.
\end{definition}
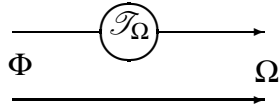
\begin{figure}[hbt]
    \setlength{\unitlength}{800sp}
    \begin{picture}(8745,3219)(931,-3565)
      {\thicklines \put(5401,-1261){\oval(1756,1756)}}
      {\put(1801,-1261){\line(1, 0){2700}}}
      {\put(6301,-1261){\vector(1, 0){3300}}}
      {\put(1801,-3361){\vector(1, 0){7800}}}
      \put(2026,-2611){\makebox(0,0)[b]{$\Phi$}}
      \put(5401,-1486){\makebox(0,0)[b]{$\tT_\Omega$}}
      \put(9676,-2811){\makebox(0,0)[b]{$\Omega$}}
    \end{picture}
\caption{Illustration of the notion of dynamically faithful state for a bipartite system (Definition \ref{prepfaith}).}
  \end{figure}
Clearly a bipartite state $\Phi$ that is preparationally faithful for system 1 is also locally
preparationally faithful for system 1, namely every local state $\omega$ of system 2 can be achieved
by a suitable local transformation $\tT_\omega$ on system 1.
\medskip
\par In Postulate \ref{p:faith} we also use the notion of {\em symmetric} joint state, defined as follows.
\begin{definition}[Symmetric joint state of two identical systems] We call a joint state of two
  identical systems {\em symmetric} if for any couple of transformations $\tA$ and $\tB$ one has
\begin{equation}
\Phi(\tA,\tB)=\Phi(\tB,\tA).
\end{equation}
\end{definition}
\section{The complex Hilbert space structure for finite dimensions}\label{s:finiteHilbert}
In this section I will derive the complex Hilbert space formulation of Quantum Mechanics for finite
dimensions from the five Postulates. This will be done as follows. From Postulates \ref{p:locobs}
and \ref{p:Bell} I obtain an identity between the affine dimension of the convex set of states and
its informational dimension, corresponding to assess that the dimension of the linear space of
effects is the square of an integer number.  Then from the bilinear symmetric form over effects
given by a faithful state---whose existence is postulated in Postulate \ref{p:faith}---I derive a
strictly positive real scalar product over generalized effects, which makes their linear space a
real Hilbert space.  Finally, since the dimension of such Hilbert space is the square of an integer,
one deduces that the Hilbert space of generalized effects is isomorphic to a real Hilbert space of
Hermitian complex matrices representing selfadjoint operators over a complex Hilbert space, which is
the Hilbert space formulation of Quantum Mechanics.
\subsection{Dimensionality theorems}
We now consider the consequences of Postulates \ref{p:locobs} and \ref{p:Bell}. We will see that
they entail dimensionality theorems that agree with the tensor product rule for Hilbert spaces for
composition of independent systems in Quantum Mechanics. Moreover, Postulate \ref{p:Bell}, in
particular, will have as a consequence that generalized effects can be represented as Hermitian
complex matrices over a complex Hilbert space $\sH$ of dimensions equal to $\idim{\Stset}$, which is
the Hilbert space formulation of Quantum Mechanics.

\par The {\em local observability principle} (Postulate \ref{p:locobs}) is operationally crucial, since it
reduces enormously the experimental complexity, by guaranteeing that only local (although jointly
executed!) experiments are sufficient to retrieve a complete information of a composite system,
including all correlations between the components. The principle reconciles holism with
reductionism, in the sense that we can observe an holistic nature in a reductionistic way---\ie locally.
This principle implies the following identity for the affine dimension of a composed system
\begin{equation}\label{admbound1}
\adm(\Stset_{12})=\adm(\Stset_1)\adm(\Stset_2)+\adm(\Stset_1)+\adm(\Stset_2).
\end{equation}
We can first prove that the left side of Eq. (\ref{admbound1}) is a lower bound for the right side.
Indeed, the number of outcomes $N$ of a minimal informationally complete observable is given by
$N=\adm(\Stset)+1$, since it equals the dimension of the affine space embedding the convex set of
states $\Stset$ plus an additional dimension for normalization. Now, consider a global
informationally complete measurement made of two local minimal informationally complete observables
measured jointly. It has number of outcomes $[\adm(\Stset_1)+1][\adm(\Stset_2)+1]$.  However, we are
not guaranteed that the joint observable is itself minimal, whence the bound.  The opposite
inequality can be easily proved by considering that a global informationally incomplete measurement
made of minimal local informationally complete measurements should belong to the linear span of a
minimal global informationally complete measurement.
\par It is worth noticing that identity (\ref{admbound1}) is the same that we have in Quantum Mechanics
for a bipartite system, due to the tensor product structure.  Therefore, the tensor product is not a
consequence of dynamical independence in Def.  \ref{p:independent}, but follows from the local
observability principle.  \bigskip

Postulate \ref{p:Bell} now gives a bound for the informational dimension of the convex sets of
states. In fact, if for any bipartite system made of two identical components and for some
preparations of one component there exists a discriminating observable that is informationally
complete for the other component, this means that $\adm(\Stset)\geq\idim{\Stset^{\times 2}}-1$, with
the equal sign if the informationally complete observable is also minimal, namely
\begin{equation}\label{infcomfromdiscr}
\adm(\Stset)=\idim{\Stset^{\times 2}}-1.
\end{equation}
By comparing this with the affine dimension of the bipartite system, we get
\begin{equation}
\begin{split}
\adm(\Stset^{\times 2})=&\adm(\Stset)[\adm(\Stset)+2]=[\idim{\Stset^{\times 2}}-1][\idim{\Stset^{\times 2}}+1]\\=&\idim{\Stset^{\times 2}}^2-1,
\end{split}
\end{equation}
which, generalizing to any convex set, gives the identification
\begin{equation}\label{Hd}
\adm(\Stset)=\idim{\Stset}^2-1,
\end{equation}
corresponding to the dimension of the quantum convex sets $\Stset$ due to the underlying Hilbert
space.  Moreover, upon substituting Eq. (\ref{infcomfromdiscr}) into Eq. (\ref{Hd}) one obtain
\begin{equation}\label{Hd2}
\idim{\Stset^{\times 2}}=\idim{\Stset}^2,
\end{equation}
which is the quantum product rule for informational dimensionalities corresponding to the quantum
{\em tensor product}. To summarize, it is worth noticing that the {\em quantum dimensionality rules}
(\ref{Hd}) and (\ref{Hd2}) follow from Postulates \ref{p:locobs} and \ref{p:Bell}.
\par To conclude this section we notice that Postulate \ref{p:faith} immediately implies the following identity
\begin{equation}
\adm(\Trnset)=\adm(\Stset^{\times 2})+1.
\end{equation}
\subsection{Derivation of the complex Hilbert space structure}
The faithful state $\Phi$ naturally provides a bilinear form $\Phi(\cA,\cB)$ over effects $\cA,\cB$,
which is certainly positive over physical effects, since $\Phi(\cA,\cA)$ is a probability.  However,
unfortunately, the fact that the form is positive over physical effects doesn't guarantee that it
remains positive when extended to the linear space of generalized effects, namely to their linear
combinations with real (generally non positive) coefficients. This problem can be easily cured by
considering the absolute value of the bilinear form $|\Phi|:=\Phi_+-\Phi_-$, and then adopting
$|\Phi|(\cA,\cB)$ as the definition for the scalar product between $\cA$ and $\cB$.  The absolute
value $|\Phi|$ can be defined thanks to the fact that $\Phi$ is real symmetric, whence it can be
diagonalized over the linear space of generalized effects. Upon denoting by $\map{P}_\pm$ the
orthogonal projectors over the linear space corresponding to positive and negative eigenvalues,
respectively, one has $\Phi_\pm=\Phi(\cdot,\map{P}_\pm\cdot)$, namely
\begin{equation}\label{scalprod1}
|\Phi|(\cA,\cB)=\Phi(\cA,\varsigma(\cB)),\quad\varsigma(\cA)=(\map{P}_+-\map{P}_-)(\cA).
\end{equation}
The map $\varsigma$ is an involution, namely $\varsigma^2=\map{I}$. Notice that there is no non zero
generalized effect $\cC$ with $|\Phi|(\cC,\cC)=0$. Indeed, the requirement that the state $\Phi$ is
also preparationally faithful implies that for every state $\omega$ there exists a suitable
transformation $\tT_\omega$ such that $\omega= \Phi_{\tI,\tT_\omega}|_1$
with $\Phi(\tI,\tT_\omega)>0$, whence
\begin{equation}\label{scalprod2}
\omega(\cC)=\Phi_{\tI,\tT_\omega}|_1(\cC)=\Phi(\cC,\varsigma(\widetilde\cT_\omega))=
|\Phi|(\cC,\widetilde\cT_\omega),\qquad \widetilde\cT_\omega=\frac{\varsigma(\cT_\omega)}{\Phi(\cI,\cT_\omega)}, 
\end{equation}
and due to non-negativity of $|\Phi|$ one has
\begin{equation}\label{smartbound}
\omega(\cC)\leq\sqrt{|\Phi|(\cC,\cC)\;
|\Phi|(\widetilde\cT_\omega,\widetilde\cT_\omega)},
\end{equation}
which implies that $\omega(\cC)=0$ for all states $\omega$, \ie $\cC=0$.  Therefore,
$|\Phi|(\cA,\cB)$ defines a strictly positive real symmetric scalar product, whence the linear space
$\Cntset_\Reals$ of generalized effects becomes a real pre-Hilbert space. The Hilbert space is then
obtained by completion in the norm topology (for the operational relevance of norm closure see
Remark \ref{r:closure}), and we will denote it by $\sW_\Phi$.  Notice that $\sW_\Phi$ is a real
Hilbert space, since both its linear space and the scalar product are real. For finite dimensional
convex set $\Stset$ one has
\begin{equation}\label{dimWset}
\dim(\sW_\Phi)=\adm(\Stset)+1,
\end{equation}
since the linear space of generalized effects $\Cntset_\Reals$ is just the space of the linear
functionals over $\Stset$, with one additional dimension corresponding to normalization. But from
Eqs. (\ref{Hd}) and (\ref{dimWset}) it follows that
\begin{equation}\label{dimWset2}
\dim(\sW_\Phi)=\idim{\Stset}^2.
\end{equation}
The last identity implies that the real Hilbert space $\sW_\Phi$ is isomorphic to the real Hilbert
space of Hermitian complex matrices representing selfadjoint operators over a complex Hilbert space
$\sH$ of dimensions $\dim(\sH)=\idim{\Stset}$: this is the Hilbert space formulation of Quantum
Mechanics.  Indeed, this is sufficient to recover the full mathematical structure of Quantum
Mechanics, since once the generalized effects are represented by Hermitian matrices, the physical
effects will be represented as elements of the truncated convex cone of positive matrices, the
physical transformations will be represented as CP identity-decreasing maps over effects, and
finally, states will be represented as density matrices via the Bush version \cite{busch} of the
Gleason theorem, or via our state-effect correspondence coming from the preparationally faithfulness
of $\Phi$.
\section{Infinite dimensions: the C${}^*$-algebra of generalized transformations}
In the previous section I derived the Hilbert space formulation of Quantum Mechanics in the finite
dimensional case. Such derivation does not hold for infinite dimension, since we cannot rely on the
dimensionality identities proved in Section \ref{s:finiteHilbert}. In the infinite dimensional case
we need an alternative way to derive Quantum Mechanics, such as the construction of a C$^*$-algebra
representation of generalized transformations. In order to do that we need to extend the real Banach
algebra $\Trnset_\Reals$ to a complex algebra, and for this we need to derive the {\em adjoint} of a
transformation from the five postulates: this is the goal of the present section. It will turn out
that only four of the five postulates are now needed. The adjoint is given as the composition of
{\em transposition} and {\em complex-conjugation} of physical transformations, both maps being
introduced operatonally on the basis of the existence of a symmetric dynamically faithful state due
to Postulate \ref{p:faith}.  The {\em complex conjugate} map will be an extension to
$\Trnset_\Reals$ of the involution $\varsigma$ of Section \ref{s:finiteHilbert}. With such an
adjoint I will then derive a Gelfand-Naimark-Segal (GNS) representation \cite{GelfandNeumark} for
transformations, leading to a C${}^*$-algebra.
\subsection{The transposed transformation}
\par For a symmetric bipartite state that is faithful both dynamically and preparationally, for every
transformation on system 1 there always exists a (generalized) transformation on system 2 giving the
same operation on that state. This allows us to introduce operationally the following notion of {\em
  transposed transformation}.
\begin{definition}[Transposed transformation]
  For a {\em faithful} bipartite state $\Phi$, the {\em transposed transformation} $\tA'$ of the
  transformation $\tA$ is the generalized transformation which when applied to the second component
  system gives the same conditioned state and with the same probability as the transformation
  $\tA$ operating on the first system, namely
\begin{equation}\label{twinid}
(\tA,\tI)\Phi=(\tI,\tA')\Phi
\end{equation}
\end{definition}
\bigskip
\begin{figure}[hbt]
 \setlength{\unitlength}{800sp}
    \begin{picture}(8745,3219)(5931,-3565)
      {\thicklines \put(5401,-1261){\oval(1756,1756)}}
      {\put(1801,-1261){\line(1, 0){2700}}}
      {\put(6301,-1261){\vector(1, 0){3300}}}
      {\put(1801,-3361){\vector(1, 0){7800}}}
      \put(2026,-2611){\makebox(0,0)[b]{$\Phi$}}
      \put(5401,-1486){\makebox(0,0)[b]{$\tA$}}
      \put(9676,-2811){\makebox(0,0)[b]{$(\tA,\tI)\Phi$}}
    \end{picture}
\hskip 1truecm
  \setlength{\unitlength}{800sp}
    \begin{picture}(8745,3219)(3931,-3565)
      {\thicklines \put(5401,-3461){\oval(1756,1756)}}
      {\put(1801,-3361){\line(1, 0){2700}}}
      {\put(6301,-3361){\vector(1, 0){3300}}}
      {\put(1801,-1261){\vector(1, 0){7800}}}
      \put(2026,-2611){\makebox(0,0)[b]{$\Phi$}}
      \put(5401,-3686){\makebox(0,0)[b]{$\tA'$}}
      \put(11676,-2811){\makebox(0,0)[b]{$(\tI,\tA')\Phi\equiv(\tA,\tI)\Phi$}}
    \end{picture}
\caption{\small Illustration of the operational concept of {\em transposed transformation}.}
\end{figure}
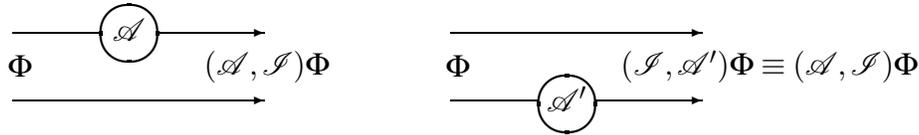
\bigskip
Eq. (\ref{twinid}) is equivalent to the following identity
\begin{equation}
\Phi(\tB\circ\tA,\tC)=\Phi(\tB,\tC\circ\tA').
\end{equation}
Clearly one has $\tI'=\tI.$ It is easy to check that $\tA\to\tA'$ satisfies the axioms of transposition
\begin{equation}\label{d:genad}
1. \;(\tA+\tB)'= \tA'+\tB',\qquad 2.\; (\tA')'=\tA,\qquad 3.\; (\tA\circ\tB)'= \tB'\circ\tA'.
\end{equation}
Indeed, axiom 1 is trivially satisfied, whereas axiom 2 is proved as follows
\begin{equation}\label{ax1}
\begin{split}
\Phi(\tB\circ\tA'',\tC)=&\Phi(\tB,\tC\circ\tA')=\Phi(\tC\circ\tA',\tB)=\Phi(\tC,\tB\circ\tA)\\ =&
\Phi(\tB\circ\tA,\tC),
\end{split}
\end{equation}
and, finally, for axiom 3 one has
\begin{equation}\label{ax3}
\Phi(\tC\circ(\tB\circ\tA),\tD)=\Phi(\tC\circ\tB,\tD\circ\tA')=\Phi(\tC,\tD\circ\tA'\circ\tB'),
\end{equation}
whereas unicity is implied by faithfulness.
\subsection{The complex conjugated transformation}
Unfortunately, even though the transposition defined in identity (\ref{twinid})
works as an adjoint for the symmetric bilinear form $\Phi$ as in Eqs. (\ref{ax1}) and (\ref{ax3}),
however, it is not the right adjoint for the scalar product given by the strictly positive bilinear
form $|\Phi|(\cA,\cB)$ in Eq.  (\ref{scalprod1}), due to the presence of the involution $\varsigma$.
In order to introduce an adjoint for generalized transformations (with respect to the scalar product
between effects) one needs to extend the involution $\varsigma$ to generalized transformations. This
can be easily done, since the bilinear form of the faithful state is already defined over
generalized transformations, and $\Phi$ is symmetric over the linear space $\Trnset_\Reals$.
Therefore, with a procedure analogous to that used for effects we introduce the absolute value
$|\Phi|$ of the symmetric bilinear form $\Phi$ over $\Trnset_\Reals$, whence extend the scalar
product to $\Trnset_\Reals$.  Clearly, since the bilinear form $\Phi(\tA,\tB)$ will anyway depend
only on the informational equivalence classes $\cA$ and $\cB$ of the two transformations, one can
have different extensions of the involution $\varsigma$ from generalized effects to generalized
transformations, which work equally well.  One has
\begin{equation}
\varsigma(\tA)=:\tA^\varsigma\in\varsigma(\cA),
\end{equation}
with a transformation $\tA^\varsigma:=\varsigma(\tA)$ belonging to the informational class
$\varsigma(\cA)$. Clearly one has $\varsigma^2(\tA)=\varsigma(\tA^\varsigma)\in\cA$, and generally
$\varsigma^2(\tA)\neq\tA$, however, one can always consistently choose the extension such that it is
itself an involution (see also the following for the choice of the extension). The idea is now that
such an involution plays the role of the {\em complex conjugation}, such that the composition with
the transposition provides the adjoint. 

\subsection{The adjoint transformation}
Inspection of Eq. (\ref{ax3}) shows that in order to have the right adjoint of transformations with
respect to the scalar product, we need to define the scalar product via the bilinear form
$\Phi(\tA',\tB')$ over transposed transformations. Therefore, we define the scalar product between
generalized effects as follows
\begin{equation}\label{scalproddef}
{}_\Phi\!\<\cB|\cA\>_\Phi:=\Phi(\cB',\varsigma(\cA')).
\end{equation}
In the following we will equivalently write the entries of the scalar product as generalized
transformations or as generalized effects, with
${}_\Phi\!\<\tA|\tB\>_\Phi:={}_\Phi\!\<\cA|\cB\>_\Phi$, the generalized effects being the actual
vectors of the linear factor space of generalized transformations modulo informational equivalence.
Notice that one has 
${}_\Phi\!\<\tC\circ\cA|\tB\>_\Phi=\Phi(\cA'\circ\tC',\varsigma(\cB'))$, corresponding to the
operator-like form of the operation of transformations over effect $|\underline{\tC'\circ\tA}\>_\Phi=
|\tC'\circ\cA\>_\Phi$ which is the transposed version of the Heisenberg picture evolution
(\ref{operatingoneffect}). We can easily check the following steps 
\begin{equation}
\begin{split}
{}_\Phi\!\<\tC'\circ\tA|\tB\>_\Phi=&
\Phi(\tA'\circ\tC,\varsigma(\tB'))=\Phi(\tA',\varsigma(\tB')\circ\tC')\\
=&|\Phi|(\tA',\varsigma(\varsigma(\tB')\circ\tC')).
\end{split}
\end{equation}
Now, for {\em composition-preserving} involution (\ie
$\varsigma(\tB\circ\tA)=\tB^\varsigma\circ\tA^\varsigma$) one can easily verify that 
\begin{equation}
{}_\Phi\!\<\tC'\circ\tA|\tB\>_\Phi=
|\Phi|(\tA',\tB'\circ\varsigma(\tC'))=
{}_\Phi\!\<\tA|(\varsigma(\tC'))'\circ\tB\>_\Phi,
\end{equation}
namely,
\begin{equation}
{}_\Phi\!\<\varsigma(\tC')\circ\tA|\tB\>_\Phi=
{}_\Phi\!\<\tA|(\tB'\circ\varsigma^2(\tC'))'\>_\Phi=
{}_\Phi\!\<\tA|\tC\circ\tB\>_\Phi,
\end{equation}
whence $\tA^\dag:=\varsigma(\tA')$ works as an adjoint for the scalar product, namely
\begin{equation}
{}_\Phi\!\<\tC^\dag\circ\cA|\cB\>_\Phi={}_\Phi\!\<\cA|\tC\circ\cB\>_\Phi.
\end{equation}
In terms of the adjoint the scalar product can also be written as follows
\begin{equation}
{}_\Phi\!\<\tB|\tA\>_\Phi=\Phi|_2(\tA^\dag\circ\tB).
\end{equation}
The involution $\varsigma$ is composition-preserving if $\varsigma(\Trnset)=\Trnset$ namely if the
involution preserves physical transformations (this is true for an identity-preserving involution
$\varsigma(\tI)=\tI$ which is cone-preserving $\varsigma(\Trnset_\Reals^+)=\Trnset_\Reals^+$).
Indeed, for such an involution one can consider its action on transformations induced by the
involutive isomorphism $\omega\to\omega^\varsigma$ of the convex set of states $\Stset$ defined as
follows
\begin{equation}\label{defomegazeta}
\omega(\varsigma(\tA)):=\omega^\varsigma(\tA),\quad\forall\omega\in\Stset,\;\forall\tA\in\Trnset.
\end{equation}
Consistency of state-reduction $\omega_\tA\Longrightarrow\omega_\tA^\varsigma$ with the involution
on $\Stset$ corresponds to the identity
\begin{equation}\forall\omega\in\Stset,\;\forall\tA,\tB\in\Trnset,\quad
\omega_\tA^\varsigma(\tB)\equiv\omega_{\tA^\varsigma}(\tB^\varsigma)
\end{equation}
which, along with identity (\ref{defomegazeta}) is equivalent to
\begin{equation}\forall\omega\in\Stset,\;\forall\tA,\tB\in\Trnset,\quad
\omega(\varsigma(\tB\circ\tA))=\omega(\tB^\varsigma\circ\tA^\varsigma).
\end{equation}
The involution $\varsigma$ of $\Stset$ is just the inversion of the principal axes corresponding to 
negative eigenvalues of the symmetric bilinear form $\Phi$ of the faithful state in a minimal
informationally complete basis (the {\em Bloch representation} of Remark \ref{r:Bloch}: see also
Ref. \cite{dariano-losini2005}). 

\subsection{The Gelfand-Naimark-Segal (GNS) construction of the C${}^*$-algebra of generalized
  transformations} 
By taking complex linear combinations of generalized transformations and defining
$\varsigma(c\tA)=c^*\varsigma(\tA)$ for $c\in\Cmplx$, we can now extend the adjoint to complex
linear combinations of generalized transformations---that we will also call {\em complex-generalized
  transformations}, and will denote their linear space by $\Trnset_\Cmplx$. On the other hand, we
can trivially extend the the real pre-Hilbert space of generalized effects $\Cntset_\Reals$ to a
complex pre-Hilbert space $\Cntset_\Cmplx$ by just considering complex linear combinations of
generalized effects. The complex algebra $\Trnset_\Cmplx$ (that we will also denote by $\aA$) is now
complex Banach algebra space, and likewise $\Cntset_\Cmplx$ is a Banach space.

We have now a scalar product ${}_\Phi\!\<\tA|\tB\>_\Phi$ between transformations and an adjoint of
transformations with respect to such scalar product. Symmetry and positivity imply the bounding
\begin{equation}
{}_\Phi\!\<\tA|\tB\>_\Phi\leq\n{\tA}_\Phi\n{\tB}_\Phi,\label{boundscal}
\end{equation}
where we introduced the norm induced by the scalar product
\begin{equation}
\n{\tA}_\Phi^2\doteq{}_\Phi\!\<\tA|\tA\>_\Phi.\label{normGNS}
\end{equation}
The bounding (\ref{boundscal}) is obtained from positivity of ${}_\Phi\<\tA-z\tB|\tA-z\tB\>_\Phi$
for every $z\in\Cmplx$. Using the bounding (\ref{boundscal}) for the scalar product
${}_\Phi\!\<\tA'\circ\tA\circ\tX|\tX\>_\Phi$ we also see that the set $\aI\subseteq\aA$ of zero norm
elements $\tX\in\aA$ is a left ideal, \ie it is a linear subspace of $\aA$ which is stable under
multiplication by any element of $\aA$ on the left (\ie $\tX\in\aI$, $\tA\in\aA$ implies
$\tA\circ\tX\in\aI$).  The set of equivalence classes $\aA/\aI$ thus becomes a complex pre-Hilbert
space equipped with a symmetric scalar product, an element of the space being an equivalence class.
On the other hand, since $|\Phi|(\cX',\cX')=0\Longrightarrow\cX'=0\Longrightarrow\cX=0$ (we have
seen that $|\Phi|$ is a strictly positive form over generalized effects) the elements of $\aA/\aI$
are indeed the generalized effects, \ie $\aA/\aI\simeq\Cntset_\Cmplx$ as linear spaces. Therefore,
informationally equivalent transformations $\tA$ and $\tB$ correspond to the same vector , and there
exists a generalized transformation $\tX$ with $\n{\tX}_\Phi=0$ such that $\tA=\tB+\tX$, and
$\n{\cdot}_\Phi$, which is a norm on $\Cntset_\Cmplx$, will be just a semi-norm on $\aA$.  
We can define anyway a norm on transformations in a way analogous to (\ref{wernernorm}) as
\begin{equation}
\n{\tA}_\Phi:=\sup_{\cB\in\Cntset_\Cmplx,\n{\cB}_\Phi\leq 1}\n{\tA\circ\cB}_\Phi,
\end{equation}
where we remind that here we are using the transposed action of (\ref{operatingoneffect}). 
Completion of $\aA/\aI\simeq\Cntset_\Cmplx$ in the norm topology will give a Hilbert space that we
will denote by $\sH_\Phi$ (for the operational relevance of closure see Remark \ref{r:closure}).
Such completion also implies that $\Trnset_\Cmplx\simeq\aA$ is a complex C$^*$-algebra. Indeed the
fact that it is a complex Banach algebra can be proved in the same ways as in Theorem
\ref{t:transnorm}, whence it remained to be proved that the norm identity
$\n{\tA^\dag\circ\tA}=\n{\tA}^2$ holds. This is done as follows:
\begin{equation}\label{bratteli}
\begin{split}
\n{\tA}_\Phi^2=&
\sup_{\cB\in\Cntset_\Cmplx,\n{\cB}_\Phi\leq  1}{}_\Phi\!\<\tA\circ\cB|\tA\circ\cB\>_\Phi=
\sup_{\cB\in\Cntset_\Cmplx,\n{\cB}_\Phi\leq  1}{}_\Phi\!\<\cB|\tA^\dag\circ\tA\circ\cB\>_\Phi\\
\leq&\sup_{\cB\in\Cntset_\Cmplx,\n{\cB}_\Phi\leq 1}\n{\tA^\dag\circ\tA\circ\cB}_\Phi\equiv
\n{\tA^\dag\circ\tA}_\Phi\leq \n{\tA^\dag}_\Phi\n{\tA}_\Phi.
\end{split}
\end{equation}
From the last equation one gets $\n{\tA}_\Phi\leq \n{\tA^\dag}_\Phi$, and by taking the adjoint one
has $\n{\tA}_\Phi=\n{\tA^\dag}_\Phi$, from which it follows that the bound (\ref{bratteli}) gives
the desired norm identity $\n{\tA^\dag\circ\tA}=\n{\tA}^2$. The fact that $\aA$ is a
C${}^*$-algebra---whence a Banach algebra---also implies that the domain of definition of
$\pi_\Phi(\tA)$ can be easily extended to the whole $\sH_\Phi$ by continuity, due to the following
bounding between Cauchy sequences
\begin{equation}
\n{\pi_\Phi(\tA)\tX_n-\pi_\Phi(\tA)\tX_m}_\Phi=
\n{\tA\circ(\cX_n-\cX_m)}_\Phi\leq \n{\tA}_\Phi\n{\cX_n-\cX_m}_\Phi.
\end{equation}

The product in $\aA$ defines the action of $\aA$ on the vectors in $\aA/\aI$, by associating to each
element $\tA\in\aA$ the linear operator $\pi_\Phi(\tA)$ defined on the dense domain
$\aA/\aI\subseteq\sH_\Phi$ as follows
\begin{equation}
\pi_\Phi(\tA)|\cB\>_\Phi\doteq|\underline{\tA\circ\tB}\>_\Phi.
\end{equation}
One also has $|\underline{\tA\circ\tB}\>_\Phi=|\tA\circ\cB\>_\Phi$ corresponding to the
transposed version of (\ref{operatingoneffect}). 
\begin{theorem}[Born rule]
From the definition (\ref{scalproddef}) of the scalar product the Born rule rewrites in terms of the pairing
\begin{equation}
\omega(\cA)=\Phi|_2(\pi_\Phi(\omega)^\dag\pi_\Phi(\cA))\equiv
{}_\Phi\<\pi_\Phi(\cA)|\pi_\Phi(\omega)\>_\Phi
\end{equation}
with representations of effects and states given by
\begin{equation}
\pi_\Phi(\omega)=\widetilde\cT_\omega:=\frac{\cT_\omega'}{\Phi(\tI,\cT_\omega)},\quad
\pi_\Phi(\cA)=\cA'.
\end{equation}
The representation of transformations is given by
\begin{equation}
\omega(\cB\circ\tA)={}_\Phi\<\cB'|\pi_\Phi(\tA^\varsigma)|\pi_\Phi(\omega)\>_\Phi.
\end{equation}
\end{theorem}
\Proof This easily follows from the definition of preparationally faithful state. One has
\begin{equation}
\omega(\cA)=\Phi_{\tI,\tT_\omega}|_1(\cA)=
\frac{\Phi(\cA,\cT_\omega)}{\Phi(\tI,\cT_\omega)}=
|\Phi|(\cA'',\varsigma(\widetilde\cT_\omega'))=\Phi|_2(\pi_\Phi(\omega)^\dag\pi_\Phi(\cA)).
\end{equation}
For the representation of transformations one has
\begin{equation}
\begin{split}
\omega(\cB\circ\tA)=&{}_\Phi\<\pi_\Phi(\cB\circ\tA)|\pi_\Phi(\omega)\>_\Phi
={}_\Phi\<\tA'\circ\cB'|\pi_\Phi(\omega)\>_\Phi\\=&
{}_\Phi\<\cB'|\pi_\Phi(\tA^\varsigma\circ\omega)\>_\Phi=
{}_\Phi\<\cB'|\pi_\Phi(\tA^\varsigma)|\pi_\Phi(\omega)\>_\Phi.
\end{split}
\end{equation}
\qed
\section{Discussion and open problems}\label{s:openprob}
\paragraph{\bf Identity (\ref{Hd})} In deriving Eq. (\ref{Hd}) I have implicitly assumed that the
relation between the affine dimension and the informational dimension which holds for bipartite
systems must hold for any system. Indeed, assuming also that dynamically independent systems can be
made statistically independent (\ie there exist factorized states) one could independently prove that
\begin{equation}
\idim{\Stset^{\times 2}}\geq\idim{\Stset}^2,
\end{equation}
since locally perfectly discriminable states are also jointly discriminable, and the existence of a
preparationally faithful state guarantees the existence of $\idim{\Stset}^2$ jointly discriminable
states, the bound in place of the identity coming from the fact that we are not guaranteed that the
set of jointly discriminable states made of local ones is maximal. It is still not clear if the
mentioned assumption is avoidable, and, if not, how relevant it is. One may postulate that
informational laws---such as identity (\ref{Hd}) are universal, namely they are independent on the
physical system, \ie on the particular convex set of states $\Stset$. Another possibility would be to
postulate---in the spirit of experimental complexity reduction---the existence of a faithful state
which is {\em pure}: there is an hope that this will not only avoid the above mentioned
extrapolation, but also reduce the number of postulates, by dropping Postulate \ref{p:Bell}. Indeed,
neither Postulate \ref{p:Bell} nor identity (\ref{Hd}) are needed in the GNS construction for the
derivation of Quantum Mechanics in the infinite dimensional case.
\paragraph{\bf Composition-preserving involution $\varsigma$} In deriving the GNS representation of
transformations over effects we needed a composition-preserving involution $\varsigma$. As said,
composition-preserving is guaranteed if $\varsigma$ is an involution of the convex set of
states---the inversion of the principal axes corresponding to the negative eigenvalues of the
symmetric bilinear form made with the faithful state. It is still not clear if Postulates
\ref{p:independent}-\ref{p:faith} imply this. 
\bigskip
\par The above issues will be analyzed in detail in a forthcoming publication.
\section*{Appendix: Errata to Ref. \cite{dariano-losini2005} and other improvements}\label{s:erra}
The present section is given only to avoiding misunderstanding in relation to the previous work
\cite{dariano-losini2005}, and can entirely skipped by the reader.
\cite{dariano-losini2005}.
\begin{enumerate}
\item In Ref. \cite{dariano-losini2005} it was not recognized that the faithful state is generally
no longer a positive bilinear form when extended to generalized transformations/effects (although,
being a state, it is clearly positive on physical transformations/effects). This lead me to introduce
the involution $\varsigma$ in Eq. (\ref{scalprod1}) in order to define a scalar product in terms of
a positive form, with the benefit of the introduction of the adjoint.
\item In Ref. \cite{dariano-losini2005} I assumed that the transposed of a physical transformation is a
  physical transformation itself, whereas more generally one should consider it as a generalized
  transformation proportional to a physical transformation with a positive multiplication constant
  (\ie for $\tA\in\Trnset$ one has $\tA'\in\Trnset_\Reals^+$, but generally $\tA'\not\in\Trnset$).
 This was first noticed by R. Werner.
\item In Ref. \cite{dariano-losini2005} I defined the norm of generalized transformations as the norm
of generalized effects, with the result that this is only a semi-norm over transformations. Now,
using definition in Eq. (\ref{wernernorm}) the norm is strictly positive, with the benefit that the
set of generalized transformations is a Banach ${}^*$-algebra. The definition (\ref{wernernorm}) has
been suggested by R. Werner and D. Schlingeman.
\item The identity (\ref{admbound1}) was only a bound in Ref. \cite{dariano-losini2005}. The reverse
bound is now proved, based on a suggestion of P. Perinotti.
\item The stronger notion of independence used in Section \ref{s:openprob} is based on a suggestion
  of G. Chiribella and P. Perinotti.
  \item In Refs. \cite{dariano-losini2005} and \cite{darianoVax2005}) it was incorrectly argued that
  acausality of local actions is not logically entailed by system independence.
\item In Ref. \cite{dariano-losini2005} it has been incorrectly argued that every generalized
  transformation belongs to the dynamical equivalence class of a physical transformation. This is
  true only for transformations in the double cone $\Trnset_\Reals^\pm$ as now explained in Remark
  \ref{r:dcone}. This error was noticed by G. Chiribella and P. Perinotti.
\item The fact that the norm induced by the GNS construction automatically leads to a
  C${}^*$-algebra has been suggested by M. Ozawa.
\end{enumerate}
\section*{Acknowledgments}
This research has been supported by the Italian Minister of University and Research (MIUR) under
program Prin 2005.  I acknowledge very useful discussions with Reinhard Werner, Giulio Chiribella
Paolo Perinotti, Dirk Schlingeman, and Masanao Ozawa.

\end{document}